\documentclass[twocolumn,pra,amsmath,letterpaper,floatfix,showpacs]{revtex4}
\usepackage{graphicx}
\usepackage{amsmath}
\usepackage{amssymb}
\usepackage{color}

\newcommand{\Otwo}{{O$_{2}$ }}

\newcommand {\ket}[1]{|\,{#1}\,\rangle}

\begin{document}

\title{Field-free long-lived alignment of molecules in extreme rotational states}

\author{A. A. Milner$^{1}$, A. Korobenko$^{1}$, V. Milner$^{1}$}
\date{\today}

\affiliation{$^{1}$Department of  Physics \& Astronomy, The University of British Columbia, Vancouver, Canada}

\begin{abstract}
We introduce a new optical tool - a ``two-dimensional optical centrifuge'', capable of aligning molecules in extreme rotational states. Unlike the conventional centrifuge, which confines the molecules in the plane of their rotation, its two-dimensional version aligns the molecules along a well-defined axis, similarly to the effect of a single linearly polarized laser pulse, but at a much higher level of rotational excitation. The increased robustness of ultra-high rotational states with respect to collisions results in a longer life time of the created alignment in dense media, offering new possibilities for studying and utilizing aligned molecular ensembles under ambient conditions.

\end{abstract}

\maketitle

The ability to align molecules is one of the key requirements in a growing number of areas of molecular science, from attosecond high-harmonic spectroscopy\cite{Kim14,Lepine14} and photoelectron spectroscopy\cite{Stolow08} to controlling molecular interactions with atoms\cite{Tilford04}, molecules\cite{Vattuone10} and surfaces\cite{Kuipers88, Zare98, Khodorkovsky2011}, to altering molecular trajectories in external fields\cite{Purcell09, Gershnabel10} and generating THz radiation\cite{York2008, Fleischer2011}(for recent reviews on the impact of molecular alignment on molecular dynamics, see Refs.\citenum{Ohshima2010, Fleischer2012, Lemeshko2013}).

The two commonly used techniques are the adiabatic and non-adiabatic alignment with intense non-resonant laser pulses\cite{Stapelfeldt2003, Seideman2005}. Similarly to the alignment by static fields\cite{Friedrich91, Friedrich92}, adiabatic methods rely on the presence of a strong laser field during the aligned phase, which may not always be tolerated. Field-free molecular alignment is typically achieved by using the non-adiabatic interaction of a femtosecond pulse with the induced dipole moment of a molecule. Though extremely successful with cold molecular ensembles, this approach is of limited use at higher temperatures, when the required (and correspondingly higher) field strengths exceed the molecular ionization threshold.

The ionization limit can be avoided by breaking a single laser pulse into a train of pulses. Multiple schemes of using pulse trains to increase the degree of alignment have been proposed theoretically \cite{Averbukh2001, Leibscher2003, Leibscher2004, Sugny2005}, and implemented experimentally with periodic sequences of two \cite{Lee2004, Bisgaard2004, Pinkham2005} and up to eight laser pulses \cite{Cryan2009, Zhdanovich2012}. However, it has been recently shown that the centrifugal bond stretching and the effect of dynamical localization limit the reach of the non-adiabatic rotational excitation with multiple laser pulses to relatively low excitation levels\cite{Floss2013, Floss2014, Floss2015, Kamalov2015}.

Much higher degrees of rotational excitation are available through the method of an optical centrifuge\cite{Karczmarek1999, Villeneuve2000, Yuan2011, Korobenko2014a}. In contrast to the dynamics of impulsively kicked molecules, the rotation of the centrifuged molecules (known as molecular superrotors) is confined to a plane. Strong planar confinement has recently been studied experimentally by means of ion imaging\cite{Korobenko2015a} and by detecting the centrifuge-induced optical birefringence\cite{Milner2014c, Milner2015c}. Typically, the direction of the transient alignment of superrotors within the rotation plane is random from pulse to pulse, preventing one from using the centrifuge as an instrument for aligning the molecules with respect to a well-defined axis.

\begin{figure}[tb]
\includegraphics[width=1\columnwidth]{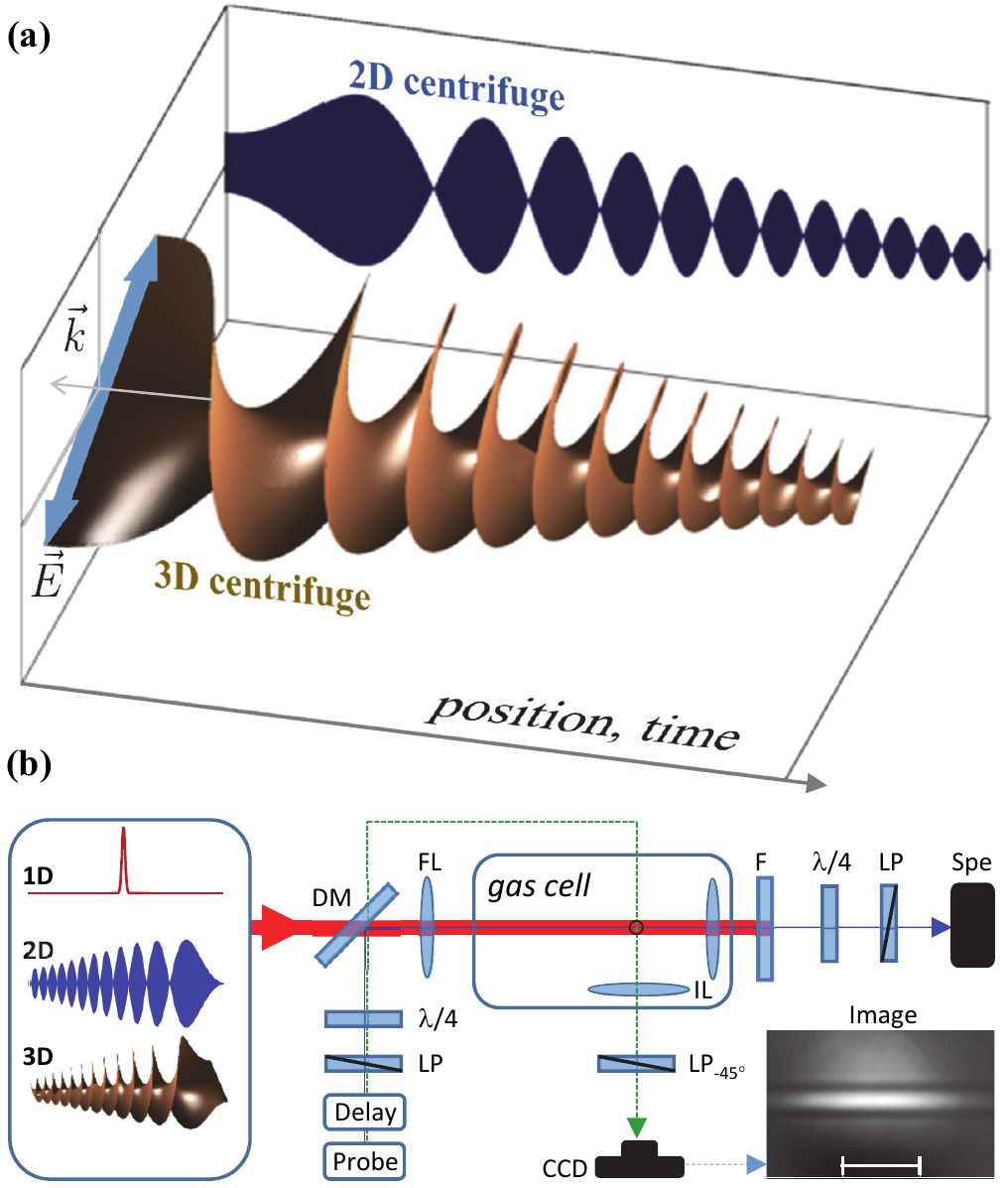}
\caption{(\textbf{a}) Illustration of the concept of a ``two-dimensional (2D) centrifuge''. The three-dimensional corkscrew-shaped surface represents the field of a conventional ``3D centrifuge'', propagating from right to left. Shown in blue is the field of a 2D centrifuge, created by passing the 3D centrifuge through a linear polarizer. (\textbf{b}) Scheme of the experimental setup. Molecules are rotationally excited by a pump beam (thick red line) focused by a focusing lens (FL) in the middle of a gas cell (black circle). Probe pulses are delayed with respect to the pump pulses, and either directed collinearly with the pump beam by means of a dichroic mirror (DM) or sent perpendicularly to it (thin blue and dashed green lines, respectively). In the longitudinal or transverse geometry, the probe light is used to image the interaction region onto an input slit of a spectrometer (Spe) or a charge-coupled device camera (CCD), respectively, with an imaging lens (IL). Linear polarizers (LP) and quarter-wave plates are used to create different combinations of linear and circular polarizers and analyzers, as described in the text. Typical image of the cloud of centrifuged molecules is shown in the lower right corner, with a horizontal bar representing the distance of 200 $\mu $m.}
\label{Fig-Setup}
\end{figure}

Here we show that a simple modification to the centrifuge allows us to achieve molecular alignment, similar to the one induced by a femtosecond kick, simultaneously with extreme levels of rotational excitation characteristic of the centrifuge spinning. We create a two-dimensional (2D) projection of the corkscrew-shaped field of a conventional centrifuge (hereafter referred to as a 3D centrifuge) by passing it through a linear polarizer. The field of a 2D centrifuge, shown in Fig.\ref{Fig-Setup}(\textbf{a}), consists of a series of pulses, all linearly polarized along the same direction. Both the individual pulse width and the time interval between the pulses are gradually decreasing from the head to the tail of the sequence.

As we demonstrate in this work, the effect of a 2D centrifuge is rather similar to that of its 3D prototype, suggesting that a ``piecewise adiabaticity''\cite{Zhdanovich2008} provides the mechanism for the accelerated rotation. The latter may be thought of as a series of Raman transitions between the states with increasing rotational quantum number, $\Delta J=2$, but constant projection of $\mathbf{J}$ on the field polarization, $\Delta M_j=0$. Since the symmetry between positive and negative values of $M_j$ is not broken (unlike the case of a 3D centrifuge, where $\Delta M_j$ is either $+2$ or $-2$) the directionality of the induced rotation is lost. In return, however, one gains the ability to fix the direction of the molecular alignment along the (now constant) field polarization.

Aligning the molecules in high $J$ states has an immediate advantage over the conventional alignment with a femtosecond pulse (hereafter referred to as a 1D kick). As has been demonstrated in our recent work, the decay of both the rotational coherence\cite{Milner2014a} and the rotational energy\cite{Milner2015c} of molecular superrotors due to collisions is much slower than that of slowly rotating molecules. Here, we compare the decay of molecular alignment, induced by a 1D kick and a 2D centrifuge in oxygen under ambient conditions, showing significantly longer decay times in the latter case.

The experimental configuration is shown in Fig.\ref{Fig-Setup}(\textbf{b}). Molecules in a gas cell are rotationally excited by either a single femtosecond pulse - a ``1D kick'', a conventional ``3D centrifuge'' or its two-dimensional projection - a ``2D centrifuge''. The excitation light (thick red line) is focused inside the gas cell to a spot size of 90 $\mu $m (FWHM diameter). For a 1D kick we use a 60 fs pulse, whose total energy of 630 $\mu $J results in the peak intensity of $5.4\times 10^{13}$ W/cm$^{2}$, close to the ionization threshold of oxygen. The 3D centrifuge is produced similarly to our previous work\cite{Korobenko2014a} and is converted to its two-dimensional version by a linear polarizer. The duration of both the 2D and 3D centrifuge pulses is about 100 ps. Given their total energy of 10 mJ (2D) and 20 mJ (3D), the respective peak intensities are $0.75\times10^{12}$ and $1.5\times10^{12}$ W/cm$^{2}$, well below the ionization threshold.

Probe pulses are 4 ps long (3.75 cm$^{-1}$ spectral bandwidth (FWHM) centered around 400 nm). In this work, we employ two different detection techniques to characterize the degree of rotational excitation and to determine the direction of the induced molecular alignment and its decay time constant. In the first approach, we send the pump and probe beams collinearly (thick red and thin blue lines in Fig.\ref{Fig-Setup}(\textbf{b}), respectively) and record the spectrum of probe pulses with a $f$/4.8 spectrometer. Owing to the induced rotational coherence, the probe spectrum develops Raman sidebands, shifted from the central probe frequency by twice the frequency of the molecular rotation. Its directionality is determined by using circularly polarized probe light as explained in detail in Ref.\citenum{Korobenko2014a}.

In the second detection method, probe pulses cross the rotationally excited molecules perpendicularly to the direction of the excitation beam (dashed green line in Fig.\ref{Fig-Setup}(\textbf{b})). We use this probe beam to image the cloud of molecular superrotors onto a CCD camera with an imaging lens (IL). The input probe polarization is linear and set at 45 degrees with respect to the vertical axis. Setting an output linear polarizer at 90 degrees to the input one, ensures that the recorded images reflect the centrifuge-induced birefringence in an otherwise isotropic oxygen gas. A typical image, taken 350 ps after the arrival of the centrifuge pulse, is shown in the lower right corner of Fig.\ref{Fig-Setup}(\textbf{b}). The image reveals a birefringent channel in the direction of the centrifuge beam, owing to the gyroscopic motion of the molecules\cite{Milner2015c}. The integral of the transmitted probe intensity, hereafter referred to as the image contrast, is used as a measure of the gas birefringence.
\begin{figure}[tb]
\includegraphics[width=.85\columnwidth]{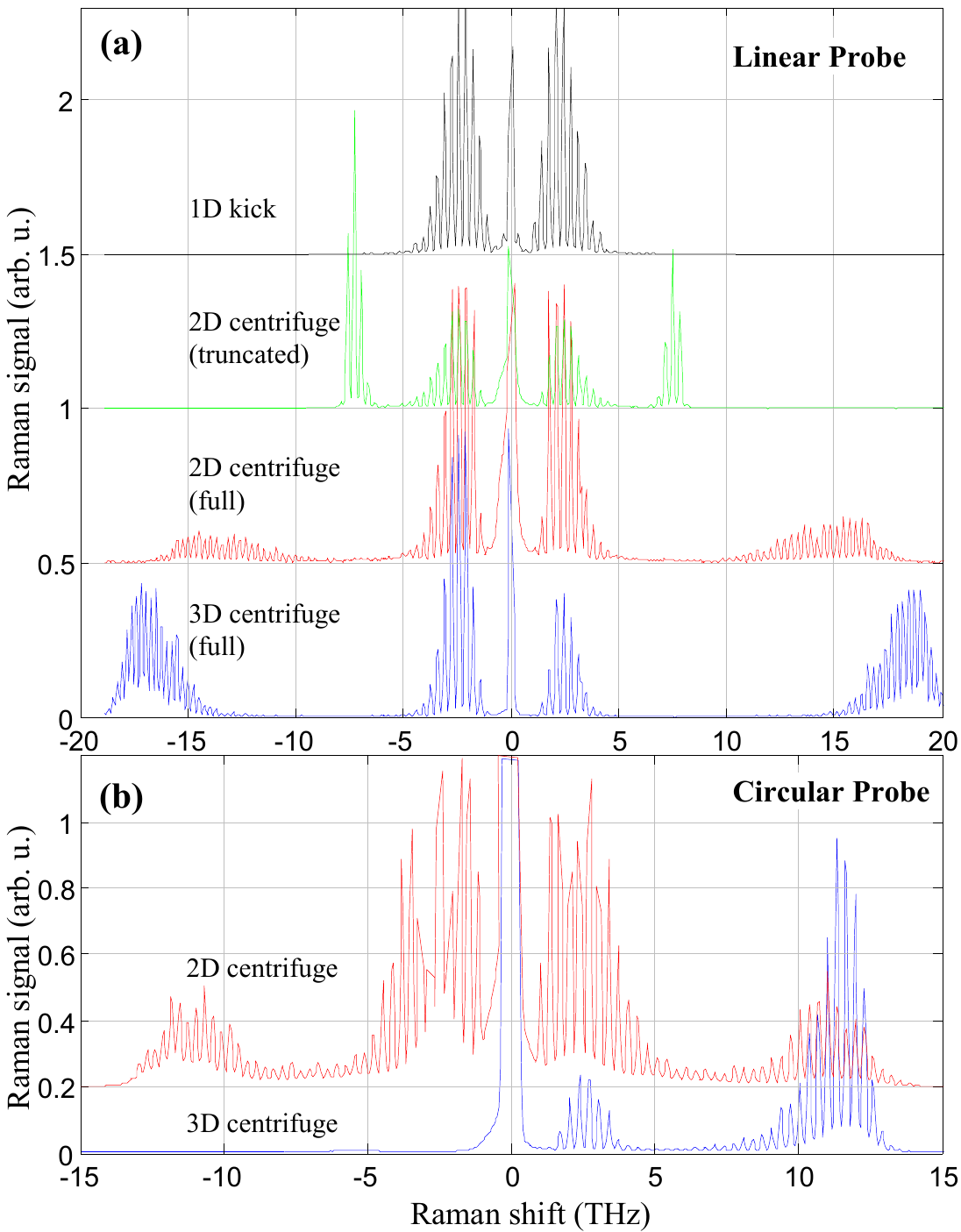}
\caption{(\textbf{a}) Rotational Raman spectra of the centrifuged oxygen molecules recorded with a linearly polarized probe light. From top to bottom, the spectra correspond to the excitation by a single femtosecond pulse (black), a ``truncated'' (see text) 2D centrifuge (green), a full 2D centrifuge (red), and a full 3D centrifuge (blue) with a terminal rotational frequency around 10 THz. The spectra have been recorded at the delay times of 400 ps. (\textbf{b}) Rotational Raman spectra of the centrifuged oxygen molecules recorded with a circularly polarized probe light. Upper red and lower blue curves correspond to the 2D and 3D centrifuge, respectively. The spectra have been recorded at the delay times of 400 ps.}
\label{Fig-Raman}
\end{figure}

To determine the frequency of the molecular rotation induced by different laser pulses, as well as its directionality, we employ coherent Raman spectroscopy, similar to our earlier work\cite{Korobenko2014a}, with pump and probe pulses collinear to one another. Raman spectrum of oxygen molecules exposed to a single intense femtosecond kick ($5.4\times 10^{13}$ W/cm$^{2}$) is plotted at the top of Fig.\ref{Fig-Raman}(\textbf{a}). The unshifted Rayleigh peak at 0 THz is surrounded by a series of Stokes and anti-Stokes lines indicative of the laser-induced coherent rotation. The amplitude envelope of these Raman lines reflects the initial thermal rotational distribution, with each line corresponding to an individual $\ket{J} \rightarrow \ket{J+2}$ (Stokes) or $\ket{J+2} \rightarrow \ket{J}$ (anti-Stokes) Raman transition.

The result of applying a conventional (3D) optical centrifuge, with its total frequency bandwidth of about 20 THz, is shown by the blue plot at the bottom of Fig.\ref{Fig-Raman}(\textbf{a}). Aside from the low-frequency thermal envelopes, corresponding to the molecules too hot to follow the adiabatic spinning\cite{Korobenko2014a}, the spectrum contains the response from the molecules centrifuged to the angular frequencies between 7.5 and 10 THz. Although the 3D centrifuge produces uni-directional molecular rotation (see below), linearly polarized probe scatters into both Stokes and anti-Stokes Raman sidebands (slightly asymmetric due to the residual ellipticity of the probe polarization).

Applying the 2D centrifuge results in the Raman spectra shown in the middle of Fig.\ref{Fig-Raman}(\textbf{a}). Even though the high-frequency Raman sidebands are somewhat lower than in the case of a 3D excitation (compare the lower red and blue curves), the 2D centrifuge is clearly producing molecular superrotors, spinning as fast as 9 THz. The appearance of a localized (in frequency) wave packet, well separated from the thermal envelope, indicates the adiabatic nature of the rotational excitation. Similar to its 3D prototype, the 2D centrifuge offers high degree of control over the frequency of molecular rotation. We illustrate this by truncating the spectral bandwidth of the pulses around 8 THz, thus moving the centrifuged rotational wave packet to lower frequencies, as shown by the green plot in Fig.\ref{Fig-Raman}(\textbf{a}).

In contrast to the uni-directional rotation of a 3D centrifuge, the 2D excitation field has no preferential sense of rotation. Owing to the selection rule $\Delta M_{j}=0$, the vectors of angular momentum of the molecular superrotors created by the 2D centrifuge lie in the plane perpendicular to its linear (and permanent) polarization, similarly to the effect of a single 1D kick. This lack of directionality is demonstrated in Fig.\ref{Fig-Raman}(\textbf{b}), where we plot Raman spectra observed with a circularly polarized probe. Missing Stokes lines in the lower (blue) spectrum indicate that the centrifuged molecules rotate in the direction opposite to the circular probe polarization\cite{Korobenko2014a}. The presence of both Stokes and anti-Stokes Raman peaks on the upper (red) plot confirms no preferential sense of rotation induced by the 2D centrifuge.
\begin{figure}[b]
\includegraphics[width=1\columnwidth]{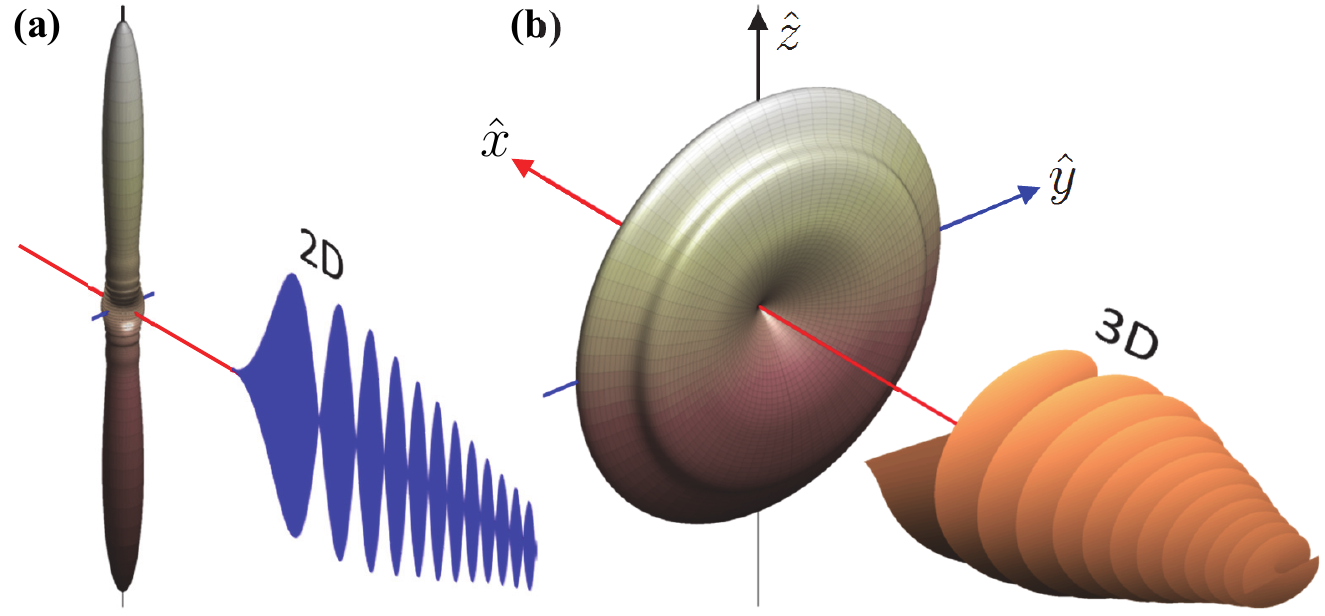}
\caption{Simulated molecular distribution after the rotational excitation by a 2D (\textbf{a}) and a 3D (\textbf{b}) optical centrifuge, propagating along the $\hat{x}$-axis. See text for details.}
\label{Fig-Distribution}
\end{figure}

To illustrate and compare the effects of the two centrifuge geometries, 2D and 3D, on the molecular alignment, we calculate the wave function of a rigid rotor adiabatically transferred from $\ket{N=1}$ (the ground rotational state of \Otwo which constitutes the majority of the centrifuged molecules) to $\ket{N=19}$, with each state split into three spin-rotational components, $J=N,N\pm1$. The relatively low rotational angular momentum of the final state is chosen for illustration purposes. Although we routinely spin oxygen to $N>100$, the calculated distributions become extremely narrow at higher values of $N$. After adding together the squared amplitudes of the spherical harmonics corresponding to the equally-populated initial $M_{j}$ states, we plot the results in Fig.\ref{Fig-Distribution}. A 2D centrifuge is seen to produce a well-aligned ensemble, shown in panel (\textbf{a}), whereas the effect of a 3D centrifuge is very different: molecules are confined in the plane of their rotation, but exhibit no preferential alignment axis, as seen in panel (\textbf{b}).
\begin{figure*}[t]
\includegraphics[width=1.99\columnwidth]{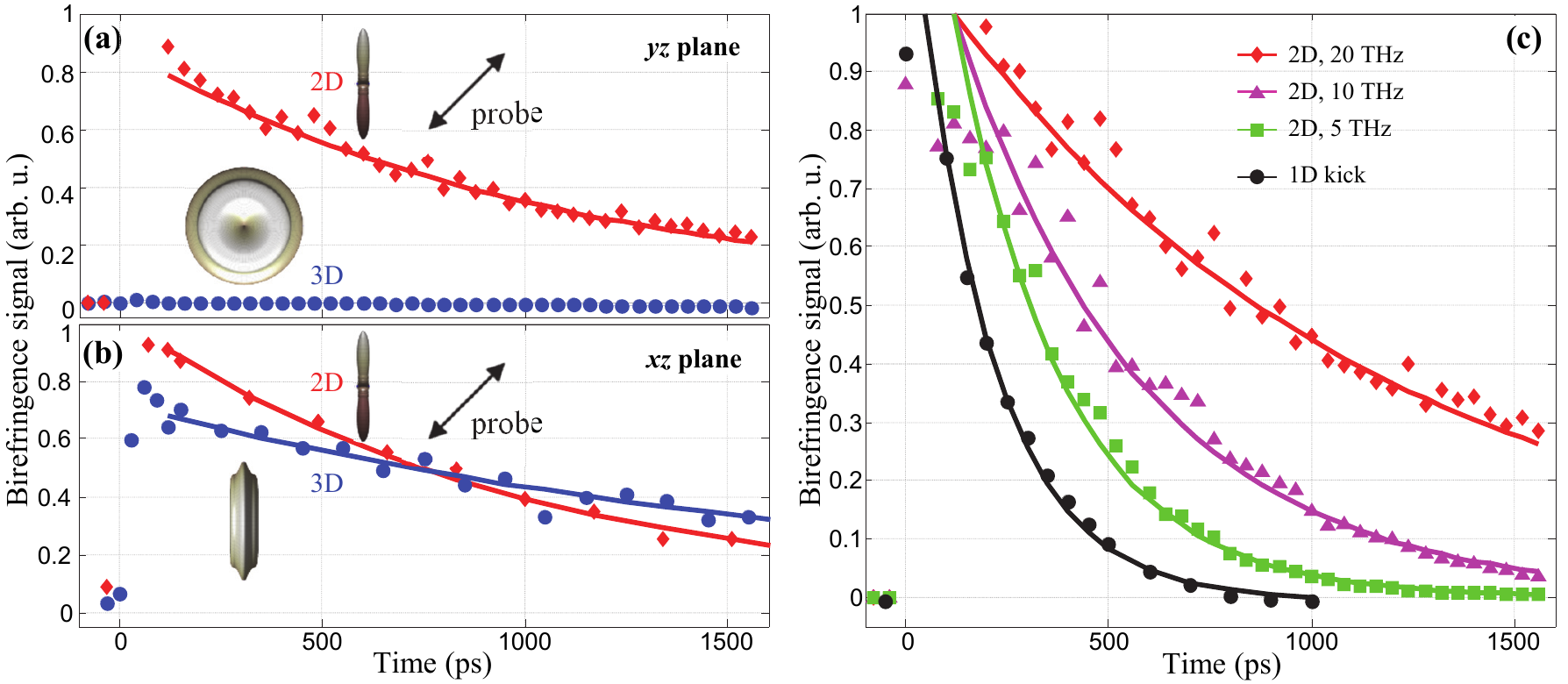}
\caption{Decay of the permanent birefringence of the rotationally excited oxygen gas in $yz$ (\textbf{a}) and $xz$ (\textbf{b}) planes. Red diamonds and blue circles correspond to the excitation by a 2D and a 3D centrifuge, respectively. Incident probe polarization, shown by the black arrow, is at 45 degrees to $\hat{z}$. The calculated distributions of molecular axes are shown to illustrate the anticipated linear anisotropy for each case. (\textbf{c}) Comparison of the birefringence decay for the 2D centrifuge with a full spectral bandwidth of 20 THz (red diamonds) and with its bandwidth truncated at $\approx 10$ THz (purple triangles) and $\lesssim 5$ THz (green squares). Black circles represent the case of a 1D kick. In all plots, solid lines show the best fit by an exponential decay.}
\label{Fig-Decay}
\end{figure*}

The shape of the molecular distributions depicted in Fig.\ref{Fig-Distribution} can be verified by measuring the permanent birefringence of the superrotors in the $xz$ and $yz$ planes. To probe the $yz$ anisotropy, we send our probe pulses collinear with the centrifuge beam (i.e. along $\hat{x}$), pass them between two crossed polarizers set at $\pm 45^\circ$ to $\hat{z}$, and record the amplitude of the Rayleigh peak in the probe spectrum as a function of the delay between the probe and centrifuge pulses. The results are shown in Fig.\ref{Fig-Decay}(\textbf{a}). No permanent birefringence is detected in the case of the 3D spinning (bottom blue circles), as one would expect for a disk-shaped distribution. On the other hand, we observe a non-zero optical birefringence induced by the 2D centrifuge, shown by red diamonds. The anisotropy axis is parallel to the linear polarization of the centrifuge pulse ($\hat{z}$), which is evident from the disappearance of the signal in the case of a vertically polarized probe (not shown), hence confirming the anticipated permanent molecular alignment of Fig.\ref{Fig-Distribution}(\textbf{a}).

To probe the anisotropy in the $xz$ plane, we switch to the transverse geometry and extract the birefringence signal from the image contrast as described earlier in the text. The 2D excitation scheme again results in the nonzero linear birefringence [red diamonds in Fig.\ref{Fig-Decay}(\textbf{b})], decaying similarly to that observed in the $yz$ plane and, therefore, confirming the rod-shaped molecular distribution. On the other hand, the disk-shaped distribution created by the 3D centrifuge, while isotropic in the $yz$ plane, exhibits an anisotropic $xz$ projection which gives rise to the optical birefringence, shown by blue circles in Fig.\ref{Fig-Decay}(\textbf{b}). We attribute the difference between the observed decay times ($970\pm41$ ps and $1850\pm138$ ps for the 2D and 3D cases, respectively) to the rather different compositions of the two created rotational wave packets. The latter can be appreciated by inspecting the corresponding Raman spectra in Fig.\ref{Fig-Raman}(\textbf{b}). Although the rotational frequency of superrotors is similar in both cases, their relative weight with respect to the low-frequency rotors is much higher in the case of the 3D centrifuge (blue curve) as compared to the 2D case (red curve). Determining the difference in the collisional cross-sections between the disk-shaped and rod-shaped distributions, which may also contribute to the unequal birefringence decay rates, requires further investigation.

Similarly to the decoherence rate of molecular superrotors\cite{Milner2014a}, the decay of the permanent molecular alignment becomes slower for faster rotating molecules due to the increased adiabaticity of collisions\cite{Khodorkovsky2015, Milner2015c}. We demonstrate this effect by comparing the decay rates of the birefringence signal observed with the 2D centrifuges which have different spectral bandwidths and, therefore, different terminal rotational frequencies. As shown in Fig.\ref{Fig-Decay}(\textbf{c}), increasing the centrifuge bandwidth from $\lesssim 5$ THz (green squares) to $\approx 8$ THz (purple triangles) and $20$ THz (red diamonds), results in the respective increasing of the exponential decay time from $269\pm9$ ps to $461\pm15$ ps and $1094\pm142$ ps.

Also plotted in Fig.\ref{Fig-Decay}(\textbf{b}) is the decay of the permanent molecular alignment induced by a single femtosecond (1D) kick (black circles). One can clearly see the noticeably shorter life time of this alignment ($187\pm8$ ps) with respect to that offered by a 2D centrifuge. Similarly to the previously discussed increasing decay rates with a slower centrifuge, the effect stems from the lower adiabaticity of collisions between the molecules exposed to a 1D kick in comparison to those excited by a 2D centrifuge.

To summarize, we proposed, implemented and studied a new method of inducing molecular alignment with intense non-resonant laser pulses. The new tool of a ``two-dimensional'' optical centrifuge has been introduced conceptually and demonstrated experimentally. We showed that the 2D centrifuge is capable of aligning molecules in extreme rotational states, i.e. creating aligned ensembles of molecular superrotors. Owing to the increased robustness of superrotors with respect to collisions, the new method offers the way of producing long-lived alignment in dense gases. Applying the 2D centrifuge to cold or $\ket{M_j=0}$ state-selected molecules will result in an extremely sharp molecular wave function with an angular width inversely proportional to the molecular angular momentum $J$. Finally, the adiabatic spinning mechanism of the centrifuge should enable transferring the molecules into a single $\ket{J>>1, M_j=0}$ state, corresponding to the truly permanent (as opposed to the time-averaged) field-free molecular alignment. The latter could be instrumental in studying multiple steric effects in molecular collisions, since in most collisional processes, the exact time of collision cannot be controlled on a femtosecond time scale.

This research has been supported by the grants from CFI, BCKDF and NSERC.

\end{document}